\documentclass[RNAAS]{aastex62}

\usepackage{amsmath}
\graphicspath{{./}{figures/}}

\received{May 30; 2020}
\accepted{July 6, 2020}

\submitjournal{RNAAS}

\begin{document}

\title{Original Research By Young Twinkle Students (ORBYTS): Ephemeris Refinement of Transiting Exoplanets II}


\author{Billy Edwards}
\affil{Department of Physics and Astronomy, University College London, London, WC1E 6BT, United Kingdom}
\affil{Blue Skies Space Ltd., 69 Wilson Street, London, EC2A 2BB, United Kingdom}

\author{Lara Anisman}
\affil{Department of Physics and Astronomy, University College London, London, WC1E 6BT, United Kingdom}

\author{Quentin Changeat}
\affil{Department of Physics and Astronomy, University College London, London, WC1E 6BT, United Kingdom}

\author{Mario Morvan}
\affil{Department of Physics and Astronomy, University College London, London, WC1E 6BT, United Kingdom}

\author{Sam Wright}
\affil{Department of Physics and Astronomy, University College London, London, WC1E 6BT, United Kingdom}

\author{Kai Hou Yip}
\affil{Department of Physics and Astronomy, University College London, London, WC1E 6BT, United Kingdom}

\author{Amiira Abdullahi}
\affil{London Academy of Excellence Stratford, Broadway House, 322 High Street, Stratford, London, E15 1AJ, United Kingdom}

\author{Jesmin Ali}	
\affil{Hammersmith Academy, 25 Cathnor Road, London, W12 9JD, United Kingdom}

\author{Clarry Amofa}	
\affil{Ark Globe Academy, Harper Road, London, SE1 6AF, United Kingdom}

\author{Antony Antoniou}	
\affil{Preston Manor School, Cartlon Avenue East, Wembley, HA9 8NA, United Kingdom}

\author{Shahad Arzouni}	
\affil{Hammersmith Academy, 25 Cathnor Road, London, W12 9JD, United Kingdom}

\author{Noeka Bradley}	
\affil{Forest School, Snaresbrook, London, E17 3PY, United Kingdom}

\author{Dayanara Campana}	
\affil{Ark Globe Academy, Harper Road, London, SE1 6AF, United Kingdom}

\author{Nandini Chavda}	
\affil{Preston Manor School, Cartlon Avenue East, Wembley, HA9 8NA, United Kingdom}

\author{Jessy Creswell}	
\affil{Ark Globe Academy, Harper Road, London, SE1 6AF, United Kingdom}

\author{Neliman Gazieva}	
\affil{Preston Manor School, Cartlon Avenue East, Wembley, HA9 8NA, United Kingdom}

\author{Emily Gudgeon-Sidelnikova}	
\affil{Hammersmith Academy, 25 Cathnor Road, London, W12 9JD, United Kingdom}

\author{Pratap Guha}	
\affil{London Academy of Excellence Stratford, Broadway House, 322 High Street, Stratford, London, E15 1AJ, United Kingdom}

\author{Ella Hayden}	
\affil{London Academy of Excellence Stratford, Broadway House, 322 High Street, Stratford, London, E15 1AJ, United Kingdom}

\author{Mohammed Huda}	
\affil{Hammersmith Academy, 25 Cathnor Road, London, W12 9JD, United Kingdom}

\author{Hana Hussein}	
\affil{Hammersmith Academy, 25 Cathnor Road, London, W12 9JD, United Kingdom}

\author{Ayub Ibrahim}	
\affil{Hammersmith Academy, 25 Cathnor Road, London, W12 9JD, United Kingdom}

\author{Chika Ike}	
\affil{Hammersmith Academy, 25 Cathnor Road, London, W12 9JD, United Kingdom}

\author{Salma Jama}	
\affil{London Academy of Excellence Stratford, Broadway House, 322 High Street, Stratford, London, E15 1AJ, United Kingdom}

\author{Bhavya Joshi}	
\affil{Preston Manor School, Cartlon Avenue East, Wembley, HA9 8NA, United Kingdom}

\author{Schet Kc}	
\affil{Preston Manor School, Cartlon Avenue East, Wembley, HA9 8NA, United Kingdom}

\author{Paris Keenan}	
\affil{Ark Globe Academy, Harper Road, London, SE1 6AF, United Kingdom}

\author{Charlie Kelly-Smith}	
\affil{Hammersmith Academy, 25 Cathnor Road, London, W12 9JD, United Kingdom}

\author{Aziza Khan}	
\affil{Preston Manor School, Cartlon Avenue East, Wembley, HA9 8NA, United Kingdom}

\author{George Korodimos}	
\affil{Hammersmith Academy, 25 Cathnor Road, London, W12 9JD, United Kingdom}

\author{Jiale Liang}
\affil{London Academy of Excellence Stratford, Broadway House, 322 High Street, Stratford, London, E15 1AJ, United Kingdom}

\author{Guilherme Luis Nogueira} 
\affil{London Academy of Excellence Stratford, Broadway House, 322 High Street, Stratford, London, E15 1AJ, United Kingdom}

\author{Neil Martey-Botchway}	
\affil{Ark Globe Academy, Harper Road, London, SE1 6AF, United Kingdom}

\author{Asan Masruri}	
\affil{Preston Manor School, Cartlon Avenue East, Wembley, HA9 8NA, United Kingdom}

\author{Osuke Miyamaru}	
\affil{Hammersmith Academy, 25 Cathnor Road, London, W12 9JD, United Kingdom}

\author{Ismail Moalin}	
\affil{Preston Manor School, Cartlon Avenue East, Wembley, HA9 8NA, United Kingdom}

\author{Fabiana Monteiro}
\affil{Ark Globe Academy, Harper Road, London, SE1 6AF, United Kingdom}

\author{Adrianna Nawrocka}	
\affil{Hammersmith Academy, 25 Cathnor Road, London, W12 9JD, United Kingdom}

\author{Sebri Musa} 
\affil{Ark Globe Academy, Harper Road, London, SE1 6AF, United Kingdom}

\author{Lilith Nelson}	
\affil{Hammersmith Academy, 25 Cathnor Road, London, W12 9JD, United Kingdom}

\author{Isabel Ogunjuyigbe}	
\affil{Hammersmith Academy, 25 Cathnor Road, London, W12 9JD, United Kingdom}

\author{Jaymit Patel}	
\affil{Preston Manor School, Cartlon Avenue East, Wembley, HA9 8NA, United Kingdom}

\author{Joesph Pereira}	
\affil{Forest School, Snaresbrook, London, E17 3PY, United Kingdom}

\author{James Ramsey}	
\affil{Hammersmith Academy, 25 Cathnor Road, London, W12 9JD, United Kingdom}

\author{Billnd Rasoul}	
\affil{Ark Globe Academy, Harper Road, London, SE1 6AF, United Kingdom}

\author{Tumo Reetsong}	
\affil{Ark Globe Academy, Harper Road, London, SE1 6AF, United Kingdom}

\author{Haad Saeed}	
\affil{Hammersmith Academy, 25 Cathnor Road, London, W12 9JD, United Kingdom}

\author{Cameron Sander}	
\affil{Forest School, Snaresbrook, London, E17 3PY, United Kingdom}

\author{Matthew Sanetra}	
\affil{Hammersmith Academy, 25 Cathnor Road, London, W12 9JD, United Kingdom}

\author{Zainab Tarabe}
\affil{Preston Manor School, Cartlon Avenue East, Wembley, HA9 8NA, United Kingdom}

\author{Milcah Tareke}
\affil{Hammersmith Academy, 25 Cathnor Road, London, W12 9JD, United Kingdom}

\author{Nazifa Tasneem}	
\affil{London Academy of Excellence Stratford, Broadway House, 322 High Street, Stratford, London, E15 1AJ, United Kingdom}

\author{Meigan Teo}	
\affil{Forest School, Snaresbrook, London, E17 3PY, United Kingdom}

\author{Asiyah Uddin}	
\affil{London Academy of Excellence Stratford, Broadway House, 322 High Street, Stratford, London, E15 1AJ, United Kingdom}

\author{Kanvi Upadhyay}	
\affil{Preston Manor School, Cartlon Avenue East, Wembley, HA9 8NA, United Kingdom}

\author{Kaloyan Yanakiev}	
\affil{Preston Manor School, Cartlon Avenue East, Wembley, HA9 8NA, United Kingdom}

\author{Deepakgiri Yatingiri}	
\affil{Preston Manor School, Cartlon Avenue East, Wembley, HA9 8NA, United Kingdom}

\author{William Dunn}
\affil{Department of Physics and Astronomy, University College London, London, WC1E 6BT, United Kingdom}

\author{Anatasia Kokori}
\affil{Department of Physics and Astronomy, University College London, London, WC1E 6BT, United Kingdom}
\affil{Birkbeck, University of London, Malet Street, London, WC1E 7HX, United Kingdom}

\author{Angelos Tsiaras}
\affil{Department of Physics and Astronomy, University College London, London, WC1E 6BT, United Kingdom}

\author{Edward Gomez}
\affil{Las Cumbres Observatory, 6740 Cortona Dr, Suite 102, Goleta, CA 93117, USA}
\affil{Cardiff University, School of Physics and Astronomy, 11-14 The Parade, Cardiff, CF24 3AA, United Kingdom}

\author{Giovanna Tinetti}
\affil{Department of Physics and Astronomy, University College London, London, WC1E 6BT, United Kingdom}
\affil{Blue Skies Space Ltd., 69 Wilson Street, London, EC2A 2BB, United Kingdom}

\author{Jonathan Tennyson}
\affil{Department of Physics and Astronomy, University College London, London, WC1E 6BT, United Kingdom}
\affil{Blue Skies Space Ltd., 69 Wilson Street, London, EC2A 2BB, United Kingdom}

\begin{abstract}

We report follow-up observations of four transiting exoplanets, TRES-2\,b, HAT-P-22\,b, HAT-P-36\,b and XO-2\,b, as part of the Original Research By Young Twinkle Students (ORBYTS) programme. These observations were taken using the Las Cumbres Observatory Global Telescope Network's (LCOGT) robotic 0.4 m telescopes and were analysed using the HOlomon Photometric Software (HOPS). Such observations are key for ensuring accurate transit times for upcoming telescopes, such as the James Webb Space Telescope (JWST), Twinkle \citep{edwards_twinkle} and Ariel \citep{tinetti_ariel}, which may seek to characterise the atmospheres of these planets. The data have been uploaded to ExoClock and a significant portion of this work has been completed by secondary school students in London.

\end{abstract}

\section{Introduction}

Transit photometry has yielded thousands of exoplanets over the last two decades. While the time for the next time is well known just after discovery, the accuracy of this prediction degrades over time to due to the accumulation of errors on the period. Unless the ephemerides are well known, upcoming facilities will be unable to characterise the atmosphere of a planet. Maintaining accurate orbital ephemerides is becoming increasingly difficult due to the sheer number of targets and will require a coordinated effort by many groups and telescope networks.

The Exoplanet Transit Database (ETD, \cite{poddany}) provides a platform for users to upload observations and citizen scientists have contributed thousands of light curves. These have been used in a number of studies \cite[e.g.][]{mallonn,edwards_orbyts}. ExoClock\footnote{\url{www.exoclock.space}} endeavours to stimulate engagement with citizen scientists who have access to telescopes, allowing them to contribute to the upcoming ESA Ariel mission. The site ranks the potential Ariel targets from \cite{edwards_ariel}, prioritising those that have a large uncertainty in their next transit time and those which have not recently been observed. These can then be filtered by the location of the observer and the telescope size, providing a list of exoplanet transits which would be observable in the near future. In collaboration with ExoClock, the Exoplanet Watch project is another coordinated effort to collect follow-up observations of exoplanet transits with small telescopes, organised in the USA. Exoplanet Watch will deliver similar information, focused on targets that will be observed by JWST \citep{zellem,zellem_exotic}.

Original Research By Young Twinkle Students (ORBYTS) is an educational programme in which secondary school pupils work on original research linked to the Twinkle Space Mission under the tuition of PhD students and other young scientists \citep{mckemmish_orbyts}. Previous projects have included calculating accurate molecular transition frequencies \citep{chubb,mckemmish_orbyts2,darby_marvel}, studying planetary aurorae \citep{wibisono2020} and spectral studies of the composition of protostellar regions \citep{holdship2019}. 

During this project, which follows on from that detailed in \cite{edwards_orbyts}, the students selected suitable follow-up targets, scheduled observations and analysed the observational data with the aim of refining planetary transit parameters.

\section{Methods}

We selected medium and high priority targets from the ExoClock website and obtained transit observations of TRES-2\,b \citep{odonovan_tres2}, HAT-P-22\,b \citep{bakos_h22}, HAT-P-36\,b \citep{bakos_h36} and XO-2\,b \citep{burke_xo2}. The observations were taken using LCOGT's network of 0.4\,m telescopes \citep{brown_lco} and analysed using HOPS\footnote{\url{https://github.com/ExoWorldsSpies/hops/}}. HOPS aligns the frames and normalises the flux of the target star by using selected comparison stars before performing a transit fit with pylightcurve \citep{tsiaras_plc}

During the fitting, the only free parameters were the planet-to-star ratio (R$_p$/R$_s$) and the transit mid time (T$_0$). The other parameters were fixed to the literature values from \cite{ozturk_tres2,bakos_h22,wang_h36,bonomo} for TRES-2\,b, HAT-P-22\,b, HAT-P-36\,b and XO-2\,b respectively and the limb darkening coefficients calculated from \cite{claretII}.

\section{Results and Future Work}

\begin{figure}
    \centering
    \includegraphics[width=0.47\columnwidth]{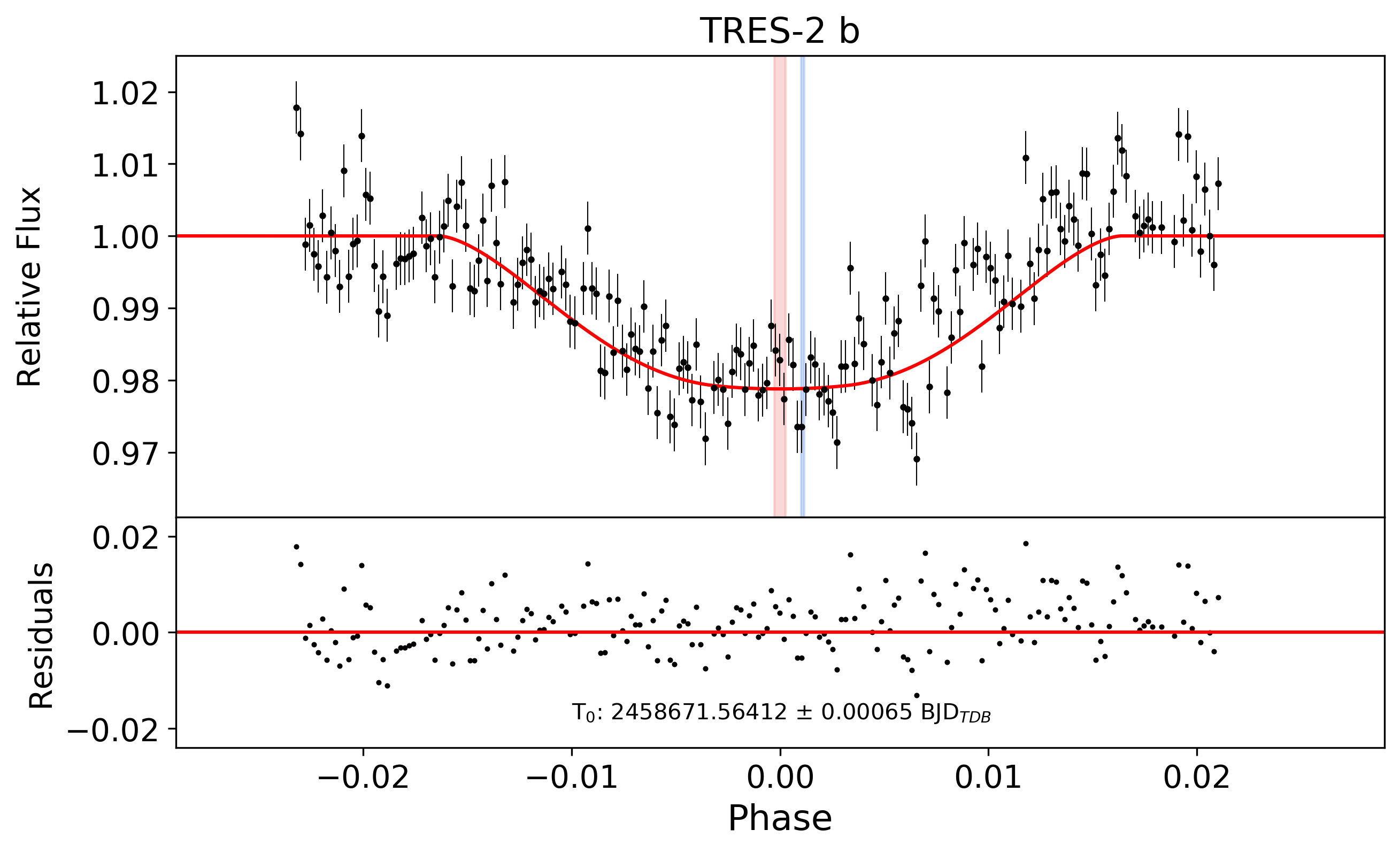}
    \includegraphics[width=0.47\columnwidth]{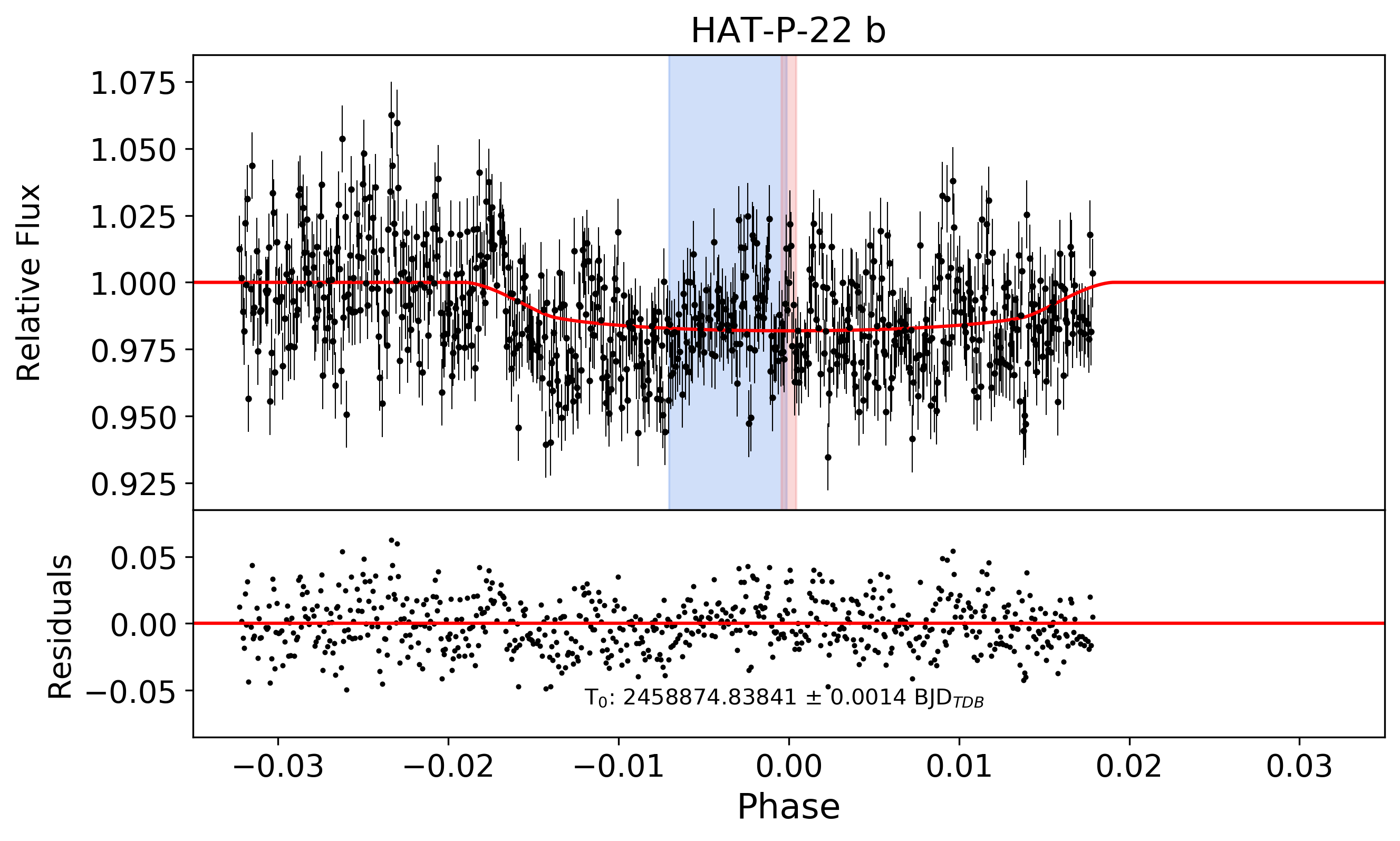}
    \includegraphics[width=0.47\columnwidth]{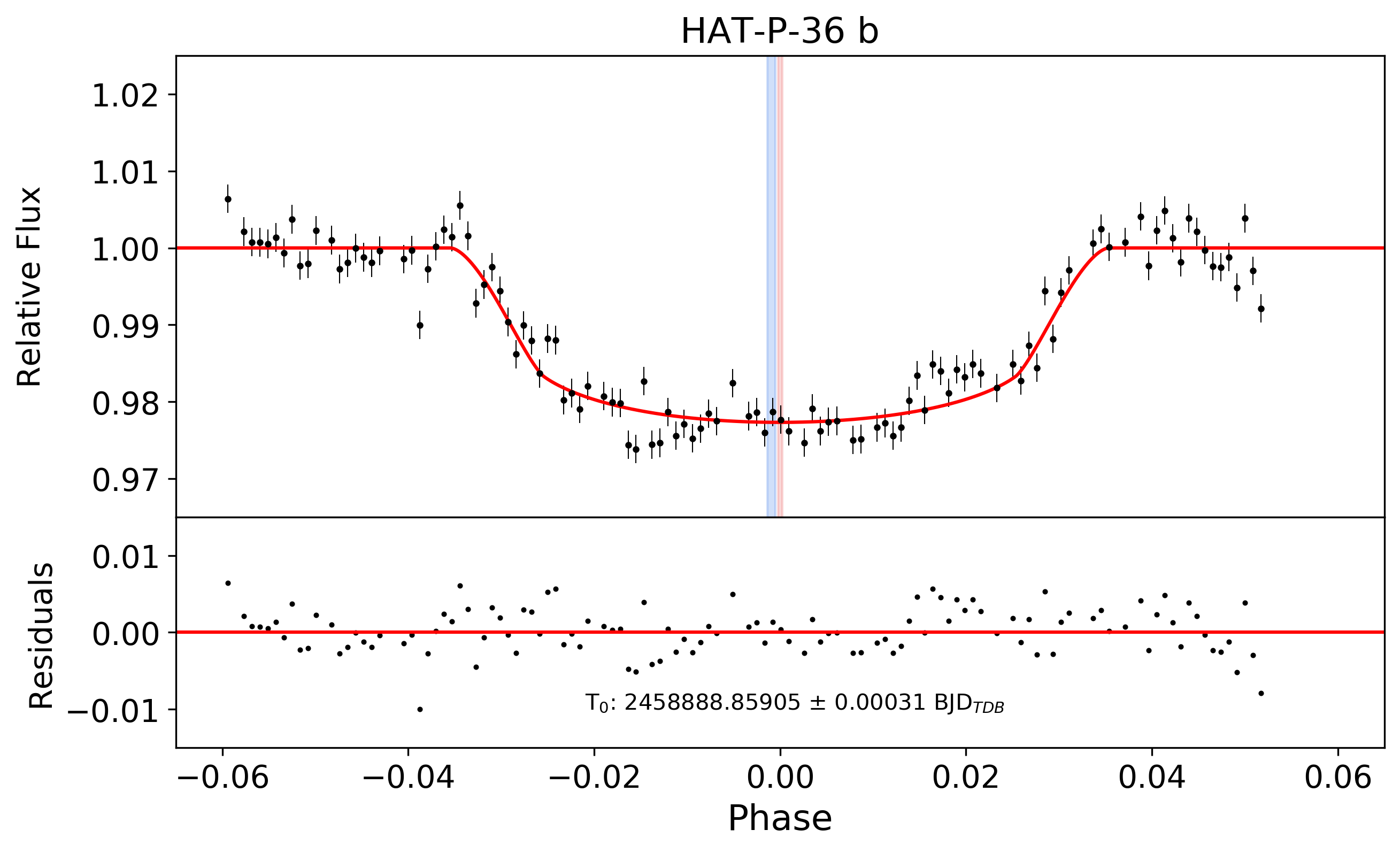}
    \includegraphics[width=0.47\columnwidth]{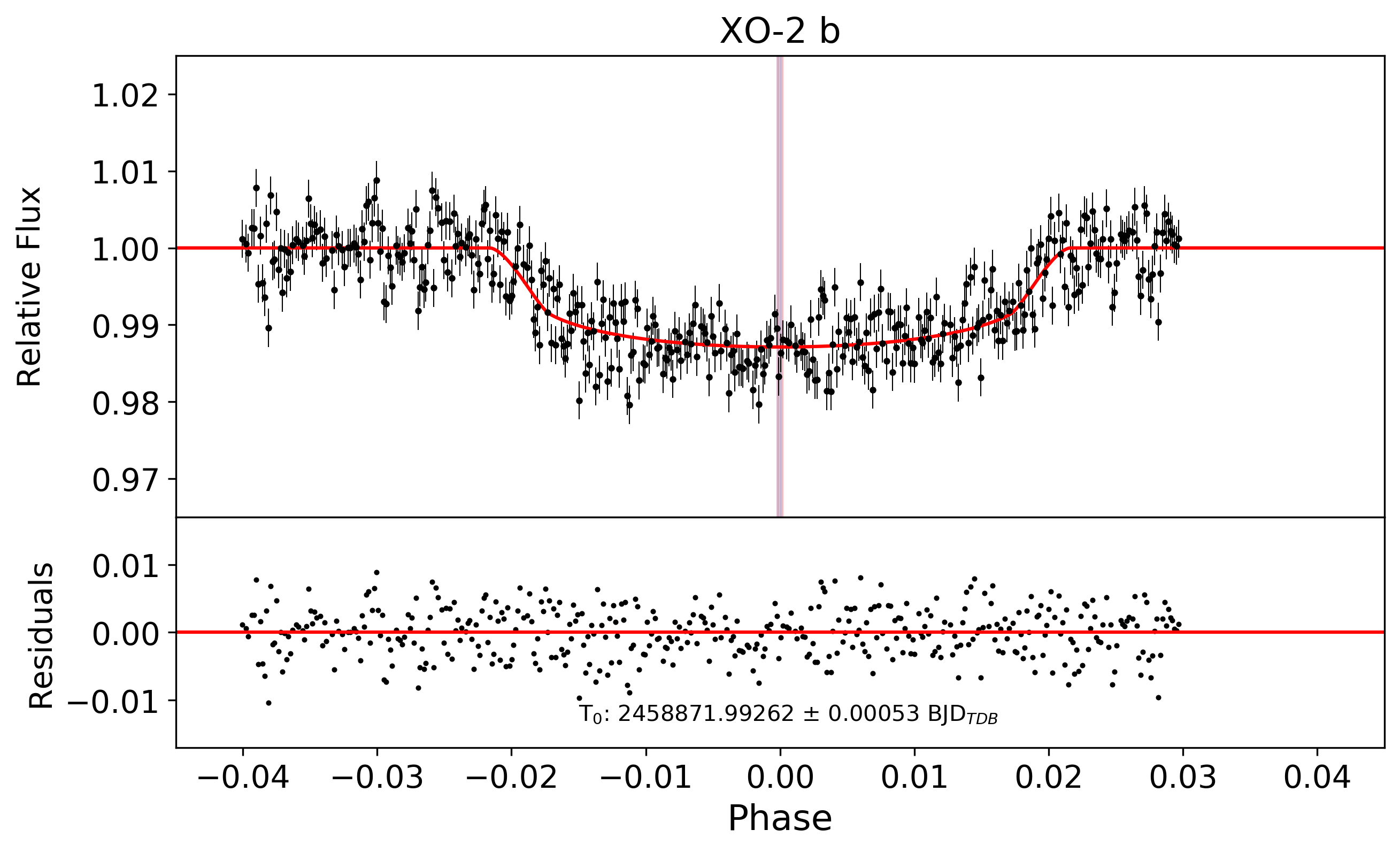}
    \caption{Transit observations obtained of TRES-2\,b, HAT-P-22\,b, HAT-P-36\,b and XO-2\,b. In each case, the data is shown in black with the best-fit transit model in red. The blue filled region indicates the predicted transit mid-time based upon literature ephemeris while the red denotes the fitted mid-time, both with 1$\sigma$ errors. The mid-times were converted to BJD$_{TDB}$ using the tool from \cite{eastman}.}
    \label{fig:lightcurves}
\end{figure}

The observed transits and recovered mid-times are shown in Figure \ref{fig:lightcurves}. For HAT-P-36\,b and XO-2\,b, the literature ephemeris predicted the mid-times both precisely and accurately, suggesting these ephemerides can be confidently used for planning future observations. ExoClock's algorithm had flagged as they had not been observed for a significant amount of time. The ephemerides of TRES-2\,b were also very precise but potentially less accurate as they do not agree with our fitted mid time. A deeper analysis, with all literature data and potentially additional observations, is needed to confirm this.

The data for HAT-P-22\,b is noticeably poorer due to the brightness of the target and because there were no other stars of a similar brightness within the telescope's field of view. The LCOGT network allows users to defocus the telescope, increasing the saturation time, and this technique would be useful for future observations. These are needed as the literature uncertainties on this planet are relatively high and, although consistent to 1$\sigma$, our observation appears to show a shift in the transit time.

Our data have been uploaded to ExoClock and, along with observations from other users, will help ensure the ephemerides of these planets are well known for potential atmospheric studies. Programmes which incorporate public engagement, educational outreach or citizen scientists have a rich literature of successful projects in the exoplanetary field \cite[e.g.][]{lintott_2013,wang_2013,baluev_2015, eisner_toi813,Nair_2020} and future ORBYTS projects will continue to collect transit data to help prepare for upcoming space-based telescopes.

\section{Acknowledgements}

BE is the PI of the LCOGT Global Sky Partners 2020 project ``Refining Exoplanet Ephemerides" and thanks the LCOGT network and its coordinators for providing telescope access. We are also thankful to the teachers at each of these schools, Phil Aspery, Matt Densham, Brook Edgar, Sarah Horn and Ehsan Pedram, for their support. BE acknowledges funding from the STFC grant ST/S002634/1.


\bibliographystyle{aasjournal}
\bibliography{main}

\end{document}